\documentclass[12pt]{article}

\usepackage{amssymb}
\usepackage{amsmath}
\usepackage{amscd}
\usepackage{latexsym}
\usepackage{graphicx}

\newcommand{\be}{\begin{equation}}
\newcommand{\ee}{\end{equation}}

\newcommand{\dlt}{\delta}
\newcommand{\prt}{\partial}
\newcommand{\br}{{\bf r}}
\newcommand{\bk}{{\bf k}}

\newcommand{\bP}{{\bf P}}

\newcommand{\bt}{\beta}

\newcommand{\ep}{\varepsilon}

\newcommand{\ra}{\rightarrow}
\newcommand{\sgm}{\sigma}

\newcommand{\gm}{\gamma}
\newcommand{\om}{\omega}

\newcommand{\dgr}{\dagger}
\newcommand{\lbd}{\lambda}
\newcommand{\Lbd}{\Lambda}

\newcommand{\rgl}{\rangle}
\newcommand{\lgl}{\langle}

\begin{document}

\begin{center}

{\Large{\bf Bose-Einstein condensation in self-consistent \\
mean-field theory} \\ [5mm]

V.I. Yukalov$^{1}$ and E.P. Yukalova$^{2}$ } \\ [3mm]

{\it
$^1$Bogolubov Laboratory of Theoretical Physics, \\
Joint Institute for Nuclear Research, Dubna 141980, Russia \\ [3mm]

$^2$Laboratory of Information Technologies, \\
Joint Institute for Nuclear Research, Dubna 141980, Russia }

\vskip 2mm

E-mail: yukalov@theor.jinr.ru

\end{center}

\vskip 0.5cm

\begin{abstract}
There is a wide-spread belief in the literature on Bose-Einstein condensation
of interacting atoms that all variants of mean-field theory incorrectly describe 
the condensation phase transition, exhibiting, instead of the necessary 
second-order transition, a first-order transition, even for weakly interacting 
Bose gas. In the present paper, it is shown that a self-consistent mean-field 
approach is the sole mean-field theory that provides the correct second-order 
condensation transition for Bose systems with atomic interactions of arbitrary 
strength, whether weak or strong.    
\end{abstract}

\newpage

\section{Introduction}

A statistical system exhibiting Bose-Einstein condensation is known to 
pertain to the universality class of the $XY$-model [1] or the two-component 
$\varphi^4$ theory [2]. Therefore, the Bose-Einstein condensation has to be 
a phase transition of second order, as has been proved for this universality 
class by renormalization-group techniques [3] and Monte Carlo simulations [4,5]. 
Bose-Einstein condensation, accompanied by superfluid transition, has been 
intensively studied in experiments with liquid $^4$He, where it is certainly 
a phase transition of second order [6]. Numerous experiments with trapped atomic 
gases also confirm the second-order of the condensation phase transition,
as is summarized in the book [7] and review articles [8-10].        

Contrary to the general requirement for the condensation transition to be 
of second order, different variants of mean-field theory, applied for 
describing this transition, lead to a first-order transition, as has been 
discussed in several articles [11-15]. The fact that such a disruption of 
the condensation phase transition to that of first order is the general 
feature of any mean-field theory was emphasized by Baym and Grinstein [16]. 
This fact also was discussed in a recent Ref. [17], concluding that none of 
the existing mean-field approaches can correctly describe the Bose-Einstein 
condensation as a second-order phase transition, leading instead to a 
first-order transition.    
 
Such a fact that no mean-field theory can reasonably describe the condensation
transition seems to be quite strange. It is a general situation that for 
practically all systems there always exists a mean-field approximation that 
provides a reasonable description of the related phase transition. Yes, it is 
known that mean-field approximations cannot yield exact critical indices, but
in the majority of cases they do correctly characterize phase transitions. Just
as one of the many examples, we can mention the magnetization transition in
the Heisenberg model, for which there exists a reasonable mean-field theory 
correctly predicting a second-order phase transition. 

It is not difficult to understand why the used mean-field theories fail in 
correctly predicting the condensation transition [11-17]. Really, the Hartree-Fock 
approximation, where the global gauge symmetry is not broken, is not able in 
principle to describe the condensed state with broken gauge symmetry, since 
the breaking of gauge symmetry is a necessary and sufficient condition 
for Bose-Einstein condensation [18,19]. The Bogolubov approximation [20,21],
by definition, is applicable only at low temperatures, where the Bose-condensed 
fraction is prevailing. The Shohno trick [22] of omitting anomalous averages
is principally not correct, as far as spontaneous gauge symmetry breaking leads 
to the simultaneous appearance of the condensate fraction as well as of these 
anomalous averages. The latter are usually of order or even larger than the 
normal averages, and in no case can be omitted [23,24]. Moreover, neglecting the 
anomalous averages renders the system unstable [9,10]. One often ascribes the 
Shohno trick of omitting anomalous averages to Popov, calling this the Popov 
approximation. It is, however, easy to check from his works [25,26] that Popov 
has never suggested such an unjustified trick. In the Girardeau-Arnowitt 
approach [27,28], there appears an unphysical gap in the spectrum of collective 
excitations. When one employs the number-conserving operator representation [29], 
one by definition is limited to very low temperatures and weak interactions, when 
almost all atoms are in the condensed state. The T-matrix approach [30], strictly 
speaking, is also applicable for low temperatures and weak interactions, but 
cannot be applied to the transition region, where, as is shown in Ref. [17], 
it leads to the system instability. More detailed discussions of these problems
can be found in Refs. [16,17,31]. 

In the majority of cases, the disruption of the phase transition to a 
discontinuous type is caused by a not self-consistent description resulting in 
the appearance of an instability caused by what Bogolubov [32,33] termed the 
{\it mismatch of approximations}. There exist two important conditions that have
to be true for a system with Bose-Einstein condensate. One condition is the 
condensate existence, formulated in one of the known forms, as is explained
in reviews [9,10], which results in the Hugenholtz-Pines relation [34] leading
to a gapless spectrum. The other is the Bogolubov-Ginibre stability condition 
[33,35] requiring that, in an equilibrium system, the condensate fraction would
minimize the thermodynamic potential. Both these conditions have to necessarily 
be treated in the same approximation. But if they are treated in different 
approximations, this immediately results in the appearance of the system 
instability. 

A novel mean-field approach has recently been advanced [36-38] satisfying the 
Hohenberg-Martin condition of self-consistency [39], being gapless and conserving.  
Employing approximate expansions in the vicinity of the critical temperature, it
has been shown that the condensation is a second-order phase transition [9,10,31,38].
In the present paper, we accomplish direct numerical calculations explicitly 
demonstrating the continuous behaviour, at the critical temperature, of the 
condensate fraction, the fraction of uncondensed atoms, the anomalous average,
sound velocity, and superfluid fraction. At the same time, the compressibility
diverges at the critical point, as it should be for a continuous transition. 
These results unambiguously prove that Bose-Einstein condensation, treated in 
the self-consistent mean-field theory, is a second-order phase transition. To our 
knowledge, this is the sole mean-field approach correctly characterizing 
Bose-Einstein condensation as a phase transition of second order.      

In Sec. 2, we briefly recall the basic points, which the self-consistent approach 
is based on. Sec. 3, presents the resulting formulas for a uniform system. 
In Sec. 4, numerical calculations for the vicinity of the critical point are 
demonstrated. Section 5 concludes.   

Throughout the paper, the system of units is employed, where the Planck and 
Boltzmann constants are set to one.

\section{Self-consistent approach}

We start with the standard energy Hamiltonian
\be
\label{1}
 \hat H = \int \hat\psi(\br) \left ( - \; \frac{\nabla^2}{2m} 
\right ) \hat\psi(\br) \; d\br \; + \;
\frac{1}{2} \; \Phi_0 
\int \hat\psi^\dgr(\br) \hat\psi^\dgr(\br) \hat\psi(\br) \hat\psi(\br) \; d\br \; ,
\ee
in which the interaction strength is given by the value
\be
\label{2}   
 \Phi_0 \equiv 4\pi \; \frac{a_s}{m} \; ,
\ee
where $a_s$ is scattering length and $m$ atomic mass. The field operators 
satisfy the Bose commutation relations. Generally, these operators depend on 
time, which is not shown for brevity.  

For describing a system with Bose-Einstein condensate, we use the Bogolubov 
shift [33] representing a field operator as the sum
\be
\label{3}
\hat\psi(\br) = \eta(\br) + \psi_1(\br)
\ee
consisting of a condensate wave function $\eta({\bf r})$ and the operator
of uncondensed atoms $\psi_1({\bf r})$. Recall that this is not an approximation, 
but an exact canonical transformation [40], so that no smallness conditions are 
imposed on the operator $\psi_1({\bf r})$, except that it satisfies the Bose
commutation relations.        

Following the Bogolubov method [32,33], the condensate function and the 
operators of uncondensed atoms are considered as separate variables that are 
orthogonal to each other in order to avoid double counting:
\be
\label{4}
 \int  \eta^*(\br) \psi_1(\br)  \; d\br = 0 \; .
\ee
The statistical average of $\psi_1({\bf r})$, representing normal atoms, is zero:
\be
\label{5}
\lgl  \psi_1(\br) \rgl = 0 \;  .
\ee
Hence the condensate function is the order parameter
\be
\label{6}
 \eta(\br) = \lgl  \hat\psi(\br) \rgl \;  .
\ee

The condensate function is normalized to the number of condensed atoms
\be
\label{7}
 N_0 = \int | \eta(\br) |^2 \; d\br \;  ,
\ee
while the number of uncondensed atoms is 
\be
\label{8}
N_1 = \int  \lgl  \psi_1^\dgr(\br) \psi_1(\br) \rgl \; d\br \;  .
\ee
Thus, the total number of atoms in the system is the sum
\be
\label{9}
  N = N_0 + N_1 \; .
\ee

The evolution equations for the variables are obtained by the extremization
of an effective action [9,10,31], which yields the equation for the condensate 
function
\be
\label{10}
i\; \frac{\prt}{\prt t} \; \eta(\br,t) = \left \lgl 
\frac{\dlt H}{\dlt \eta^*(\br,t) } \right \rgl
\ee
and the equation for the operator of uncondensed atoms
\be
\label{11}
i\; \frac{\prt}{\prt t} \; \psi_1(\br,t) = 
\frac{\dlt H}{\dlt \psi_1^\dgr(\br,t) }  \;   ,
\ee
with the grand Hamiltonian
\be
\label{12}
H = \hat H - \mu_0N_0 - \mu_1 \hat N_1 - \hat\Lbd \;   ,
\ee
in which 
\be
\label{13}
 \hat N_1 \equiv \int \psi_1^\dgr(\br)\psi_1(\br) \; d\br  
\ee
is the number-operator of uncondensed atoms and 
\be
\label{14}
 \hat\Lbd \equiv \int \left [ \lbd(\br)\psi_1^\dgr(\br) +
\lbd^*(\br) \psi_1(\br) \right ] \; d\br \;  .
\ee
The quantities $\mu_0$, $\mu_1$ and $\lambda({\bf r})$ are the Lagrange 
multipliers guaranteeing the validity of conditions (5) to (9). The 
evolution equations can be shown [31,41] to be equivalent to the Heisenberg 
equations of motion with Hamiltonian (12). For equilibrium systems, the 
statistical operator is defined [9,10,31] by the minimization of the information 
functional uniquely representing the system, which results in the operator
\be
\label{15}
\hat\rho = \frac{1}{Z} \; e^{-\bt H} \; , \qquad 
Z \equiv {\rm Tr} e^{-\bt H} \;  ,
\ee
where $\beta \equiv 1/T$ is inverse temperature. 

The average densities of condensed and uncondensed atoms, respectively, are
given by the ratios
\be
\label{16}
 \rho_0 \equiv \frac{N_0}{V} \; , \qquad \rho_1 \equiv \frac{N_1}{V} \; ,
\ee
with $V$ being the system volume and $\rho$, the total average density
\be
\label{17}
 \rho \equiv \frac{N}{V} = \rho_0 + \rho_1 \;   .
\ee

The superfluid density is defined [9,10] as 
\be
\label{18}
 \rho_s =\rho - \; \frac{{\rm var}(\hat\bP)}{3mTV} \; ,
\ee
where 
\be
\label{19}
\hat\bP \equiv \int \hat\psi(\br) ( -i{\bf\nabla} ) \hat\psi(\br) \; d\br
\ee
is the system momentum operator, whose variance is
$$
{\rm var}(\hat\bP) \equiv \lgl \hat\bP^2 \rgl - \lgl \hat\bP \rgl^2 \;  .
$$
For an equilibrium system, the total average momentum is zero, so that the 
superfluid density reduces to
\be
\label{20}
 \rho_s =\rho - \; \frac{\lgl \hat\bP^2 \rgl}{3mTV} \;  .
\ee

\section{Uniform system}

In the case of a uniform system, the condensate function is the constant
\be
\label{21}
  \eta(\br) = \sqrt{\rho_0} \; .
\ee
The operator of uncondensed atoms can be expanded over plane waves,
\be
\label{22}
 \psi_1(\br) = \frac{1}{\sqrt{V}} \; \sum_k a_k e^{i\bk\cdot\br} \;  .
\ee
Note that, because of the orthogonality condition (4), we have
$$
 \lim_{k\ra 0} a_k = 0 \;  .
$$

The operators $a_k$ in the momentum representation define the momentum 
distribution
\be
\label{23}
 n_k \equiv \lgl a_k^\dgr a_k \rgl \;   ,
\ee
called the normal average, and the anomalous average
\be
\label{24}
 \sgm_k \equiv \lgl a_k a_{-k} \rgl \;  .
\ee
Integrating distribution (23) yields the density of uncondensed atoms
\be
\label{25}
 \rho_1 = \int n_k \; \frac{d\bk}{(2\pi)^3} \;  .
\ee
Respectively, the integral of Eq. (24) gives
\be
\label{26}
\sgm_1 = \int \sgm_k \; \frac{d\bk}{(2\pi)^3} \;
\ee
that defines the density $|\sigma_1|$ of pair correlated atoms.    

Substituting into Hamiltonian (1) the Bogolubov shift (3) and using 
expansion (22), we then invoke for the operators $a_k$ the 
Hartree-Fock-Bogolubov approximation (see details in Refs. [9,10,31,37,38]).
Introducing the notation
\be
\label{27}
  \om_k \equiv \frac{k^2}{2m} + mc^2 \; ,
\ee
we find the momentum distribution
\be
\label{28}
 n_k = \frac{\om_k}{2\ep_k} \; 
\coth \left ( \frac{\ep_k}{2T} \right ) - \;
 \frac{1}{2}  
\ee
and the anomalous average
\be
\label{29}
  \sgm_k = -\;\frac{mc^2}{2\ep_k} \; 
\coth \left ( \frac{\ep_k}{2T} \right ) \;  ,
\ee
where 
\be
\label{30}
\ep_k = \sqrt{(ck)^2 + \left ( \frac{k^2}{2m} \right )^2 } 
\ee
is the spectrum of collective excitations, and the sound velocity is given 
by the equation
\be
\label{31}
mc^2 = (\rho_0 + \sgm_1) \Phi_0 \;   .
\ee
The superfluid density (20) takes the form
\be
\label{32}
 \rho_s = \rho -\; \frac{1}{3mT} 
\int k^2 \left ( n_k + n_k^2 - \sgm_k^2 \right )
\; \frac{d\bk}{(2\pi)^3} \;  .
\ee
 
Thus, for the density of uncondensed atoms (25), we have
\be
\label{33}
\rho_1 = \int \left [ \frac{\om_k}{2\ep_k} \; 
\coth \left ( \frac{\ep_k}{2T} \right ) 
- \; \frac{1}{2} \right ] \; \frac{d\bk}{(2\pi)^3} 
\ee
and for the anomalous average (26), we get
\be
\label{34}
\sgm_1 = - \int \frac{mc^2}{2\ep_k} \; 
\coth \left ( \frac{\ep_k}{2T} \right )\; \frac{d\bk}{(2\pi)^3} \; .
\ee
The superfluid density (32) becomes
\be
\label{35}
 \rho_s = \rho -\; \frac{1}{12mT} \int
\frac{k^2}{\sinh^2(\ep_k/2T)} \; \frac{d\bk}{(2\pi)^3} \;  .
\ee

It is worth stressing that the superfluid density is meaningful only if
the anomalous average (29) is taken into account. If in expression (32)
this anomalous average were omitted, then the related integral would be 
divergent leading to the meaningless value $\rho_s \ra -\infty$. The 
principal importance of the anomalous average for the correct definition 
of the superfluid density is easy to understand: The phenomenon of 
superfluidity is caused by atomic correlations. And the anomalous density
defines exactly the density of correlated atoms $|\sigma_1|$. Hence, without 
the anomalous density, there are no correlated atoms, and consequently, 
there is no superfluidity. 
    
When temperature tends to zero, then Eq. (34) leads to
\be
\label{36}
 \sgm_1 \simeq - \int \frac{mc^2}{2\ep_k} \; \frac{d\bk}{(2\pi)^3} 
\qquad 
(T \ra 0) \;  .
\ee
The latter integral is formally divergent but can be regularized, e.g., by
means of dimensional regularization that is asymptotically exact for weak
interactions [2]. So, at low temperatures and weak interactions, such that
\be
\label{37}
  \frac{T}{T_c} \ll 1 \; , \qquad \frac{\rho\Phi_0}{T_c} \ll 1 \; ,
\ee
where $T_c$ is the critical temperature, one can use the expression
\be
\label{38}
\sgm_1 \simeq - \int \frac{mc^2}{2\ep_k} \; \frac{d\bk}{(2\pi)^3} \; - \;
\int \frac{mc^2}{2\ep_k} 
\left [ \coth \left ( \frac{\ep_k}{2T} \right ) - 1 \right ] \; 
\frac{d\bk}{(2\pi)^3} \;  ,
\ee
with the appropriately regularized first term [9,10,31,37]. The behavior
of the Bose-condensed system at zero temperature, in the frame of the 
self-consistent approach, has been studied in detail in Refs. [10,37], 
exhibiting good agreement with Monte Carlo simulations [42-44]. 

The critical temperature is the temperature, where the condensate density
disappears, $\rho_0 \ra 0$. Then the gauge symmetry becomes restored, hence, 
the anomalous average also tends to zero. From the above equations, it follows 
that this happens at the temperature
\be
\label{39}
 T_c = \frac{2\pi}{m} \left [ \frac{\rho}{\zeta(3/2)} \right ]^{2/3} \;  .
\ee
In the vicinity of the critical temperature (39), Eq. (31) shows that then 
$c \ra 0$. In this critical region, the anomalous average (34) behaves as 
\be
\label{40}
 \sgm_1 \simeq -\; \frac{mc^2 T}{2\pi} \qquad (T \ra T_c ) \; .
\ee

\section{Critical region}

To study the behaviour of the system in the critical region, where $T \ra T_c$, 
it is convenient to pass to dimensionless quantities, such as the condensate
fraction $n_0$ and the fraction of uncondensed atoms $n_1$ given by the ratios
\be
\label{41}
 n_0 \equiv \frac{\rho_0}{\rho} \; , \qquad
n_1 \equiv \frac{\rho_1}{\rho} \; .
\ee
Respectively, the superfluid fraction is 
\be
\label{42}
 n_s \equiv \frac{\rho_s}{\rho} \;  .
\ee
And we introduce the dimensionless anomalous average
\be
\label{43}
 \sgm \equiv \frac{\sgm_1}{\rho} \;  .
\ee
The dimensionless sound velocity is 
\be
\label{44}
 s \equiv \frac{mc}{\rho^{1/3}} \;  .
\ee

As a dimensionless interaction strength, we use the {\it gas parameter}
\be
\label{45}
 \gm \equiv \rho^{1/3} a_s \;  .
\ee
This parameter is very natural, measuring the ratio of potential to kinetic 
energy. Really, potential energy per atom is proportional to $\rho a_s/m$,
while kinetic energy is of order $\rho^{2/3}/m$. Their ratio gives precisely 
the gas parameter (45). 

Under the validity of inequalities (37), when the second of them reads as 
$\gamma \ll 0.3$, one can use the low-temperature form (38) of the anomalous 
average. But in the critical region, one has to employ the anomalous 
average (40). 
 
Let us measure temperature in units of $\rho^{2/3}/m$, so that in what 
follows $T$ implies the dimensionless quantity, such that the transformation
to the dimensional temperature corresponds to the change
$$
T ~ \ra ~ \frac{mT}{\rho^{2/3}} \;  .
$$
Thus, for the critical temperature we have
$$
 T_c ~ \ra ~  \frac{mT_c}{\rho^{2/3}} = \frac{2\pi}{[\zeta(3/2)]^{2/3}}
= 3.312498 \; .
$$

Our aim is to investigate the behaviour of the characteristic quantities in 
the critical region close to the critical temperature $T_c$. We shall study
the behaviour of the condensate fraction
\be
\label{46}
 n_0 = 1 - n_1 \;  ,
\ee
expressed through the normal fraction 
\be
\label{47}
  n_1 = \frac{s^3}{3\pi^2} \left \{ 1 + 
\frac{3}{2\sqrt{2} } \int_0^\infty \left ( \sqrt{1+x^2}-1 \right )^{1/2}
\left [ \coth \left ( \frac{s^2 x}{2T} \right ) -1 \right ] \; dx \right \} \; ,
\ee
the dimensionless anomalous average
\be
\label{48}
\sgm = -\; \frac{sT}{2\pi} \;   ,
\ee
sound velocity $s$ given by the equation
\be
\label{49}
s^2 = 4\pi\gm (n_0 + \sgm) \;   ,
\ee
and the superfluid fraction
\be
\label{50}
 n_s = 1 - \; \frac{s^5}{6\sqrt{2}\;\pi^2 T} \int_0^\infty
\frac{(\sqrt{1+x^2}-1)^{3/2} xdx }{\sqrt{1+x^2}\;\sinh^2(s^2x/2T)} \; .
\ee
     
Solving the system of equations (46) to (50), we find that all these
characteristics of interest are continuously varying in the vicinity 
of the critical temperature. The results of numerical calculations
are presented for the condensate fraction in Fig. 1, the fraction of 
uncondensed atoms, in Fig. 2, for the anomalous average, in Fig. 3,
for the sound velocity, in Fig. 4, and for the superfluid fraction,
in Fig. 5. The continuous variation is a typical feature of the phase 
transition of second order. Let us stress that, as is seen from our 
calculations, the order of the phase transition does not depend on the 
atomic interaction strength, being of second order for any value of
the gas parameter $\gamma$. 

At the point of a second-order phase transition, the compressibility
should be divergent. The isothermal compressibility can be calculated 
by invoking different representations [45], for instance  
$$
 \kappa_T = \frac{{\rm var}(\hat N)}{\rho T N} \qquad 
( \hat N = N_0 + \hat N_1 )\;  .
$$
In the used approximation, we obtain
$$
 \kappa_T = \frac{1}{m\rho c^2} \;  .
$$
Since the sound velocity $c$ tends to zero at $T_c$, the compressibility
diverges, as it should be for a second-order phase transition.

\section{Conclusion}

We have shown that in the self-consistent mean-field theory [9,10,31,36-38]
Bose-Einstein condensation is a second-order phase transition. This is proved
by the direct numerical investigation for the critical behaviour of the most 
important quantities characterising the condensation phase transition: 
the condensate fraction, the fraction of uncondensed atoms, the anomalous 
average, the sound velocity, and the superfluid fraction. The condensate 
fraction, anomalous average, sound velocity, and the superfluid fraction 
continuously tend to zero at the same critical temperature $T_c$. Respectively,
the isothermal compressibility, inversely proportional to the sound velocity 
squared, diverges at the critical point. All this behaviour unambiguously 
demonstrates the second order of the Bose-Einstein condensation transition. 

It is worth emphasizing that all other known mean-field approximations, as is 
discussed in Refs. [11-17,31], incorrectly describe the Bose condensation as 
a first-order phase transition. In Ref. [17] it is claimed that in the 
self-consistent mean-field theory [9,10,31,36-38] the condensation transition 
is also of first order. However, this claim is caused by an error: in their 
calculations, the authors of Ref. [17] have taken the low-temperature 
approximation for the anomalous average, instead of the correct form (48) 
valid for the critical region. As we have shown in the present paper by 
straightforward numerical calculations, in the self-consistent mean-field 
theory [9,10,31,36-38] the Bose-Einstein condensation is undoubtedly a 
second-order phase transition, independently of the strength of atomic
interactions. To our knowledge, this is the sole variant of the available
mean-field approaches qualifying this transition as a second-order one.            
 
\vskip 5mm

{\bf Acknowledgment}

\vskip 3mm

Financial support from the Russian Foundation for Basic Research is acknowledged.

\newpage

\begin{center}
{\bf{\Large Figure Captions } }
\end{center}

\vskip 3cm

{\bf Figure 1}. Condensate fraction $n_0$ as a function of dimensionless temperature 
(in units of $\rho^{2/3}/m$) in the critical region.  

\vskip 1cm
{\bf Figure 2}. Fraction of uncondensed atoms $n_1$ as a function of dimensionless
temperature in the critical region.

\vskip 1cm
{\bf Figure 3}. Anomalous average $\sigma$ as a function of dimensionless temperature 
in the critical region. 

\vskip 1cm
{\bf Figure 4}. Dimensionless sound velocity $s$ as a function of dimensionless 
temperature in the critical region. 

\vskip 1cm
{\bf Figure 5}. Superfluid fraction $n_s$ as a function of dimensionless temperature 
in the critical region. 

\newpage

\begin{figure}[ht]
\centerline{\includegraphics[width=15cm]{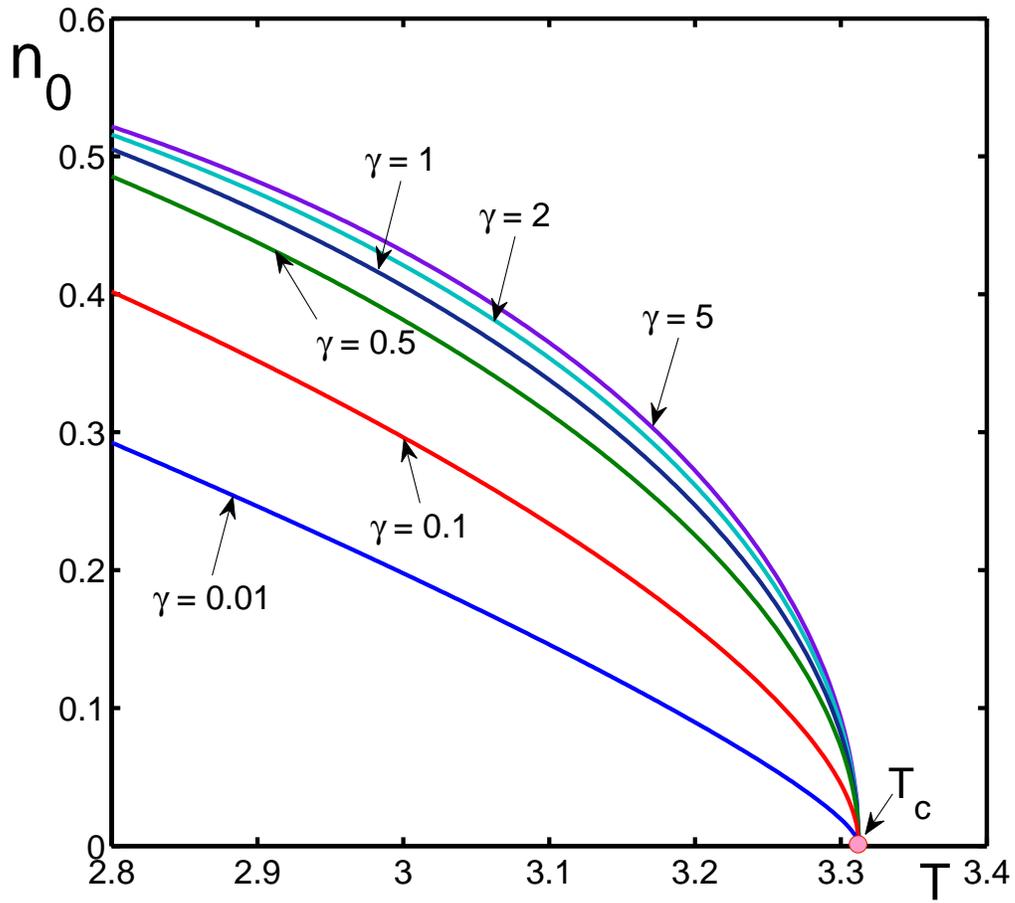} }
\caption{Condensate fraction $n_0$ as a function of dimensionless temperature 
(in units of $\rho^{2/3}/m$) in the critical region.}
\label{fig:Fig.1}
\end{figure}

\begin{figure}[ht]
\centerline{\includegraphics[width=15cm]{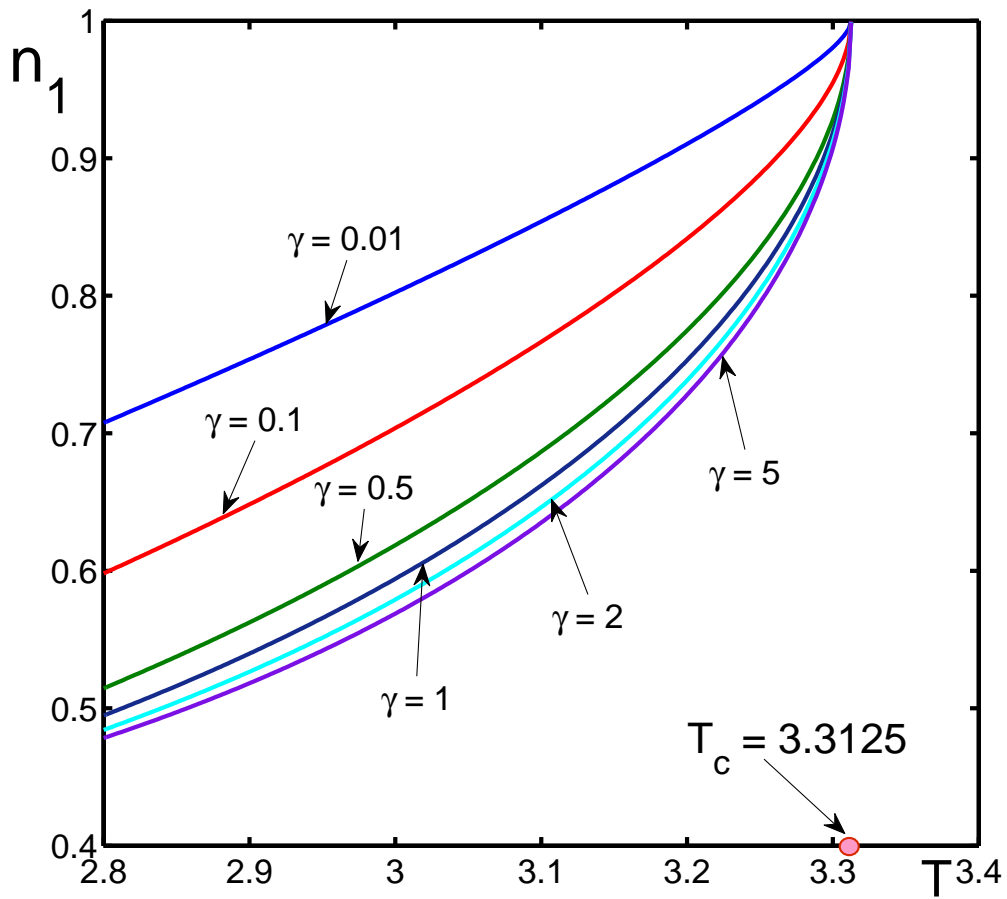} }
\caption{Fraction of uncondensed atoms $n_1$ as a function of dimensionless
temperature in the critical region.}
\label{fig:Fig.2}
\end{figure}

\begin{figure}[ht]
\centerline{\includegraphics[width=15cm]{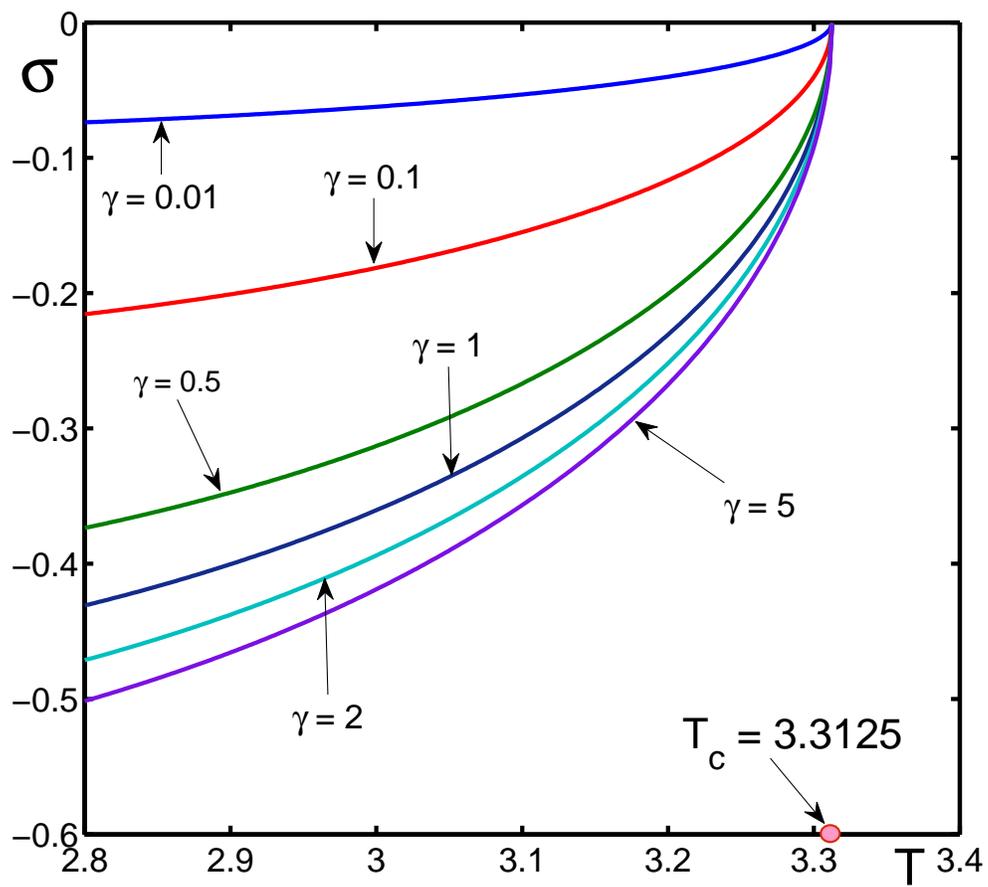} }
\caption{Anomalous average $\sigma$ as a function of dimensionless temperature 
in the critical region. }
\label{fig:Fig.3}
\end{figure}

\begin{figure}[ht]
\centerline{\includegraphics[width=15cm]{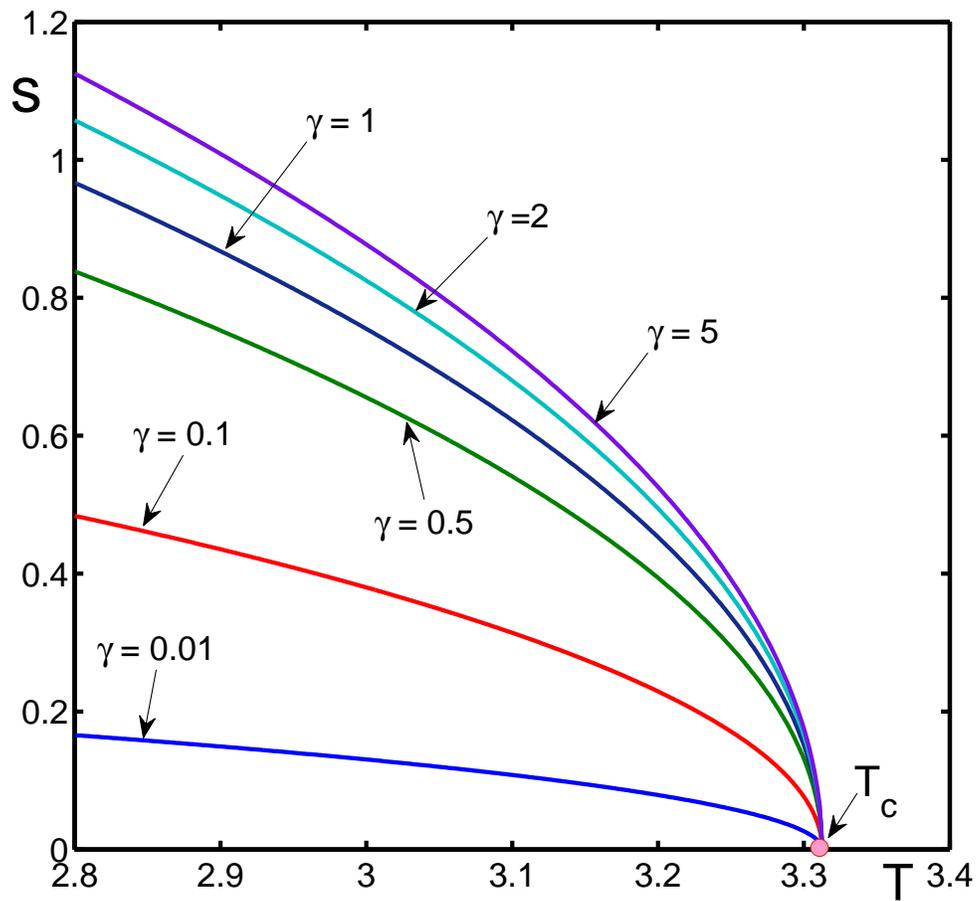} }
\caption{Dimensionless sound velocity $s$ as a function of dimensionless 
temperature in the critical region.}
\label{fig:Fig.4}
\end{figure}

\begin{figure}[ht]
\centerline{\includegraphics[width=15cm]{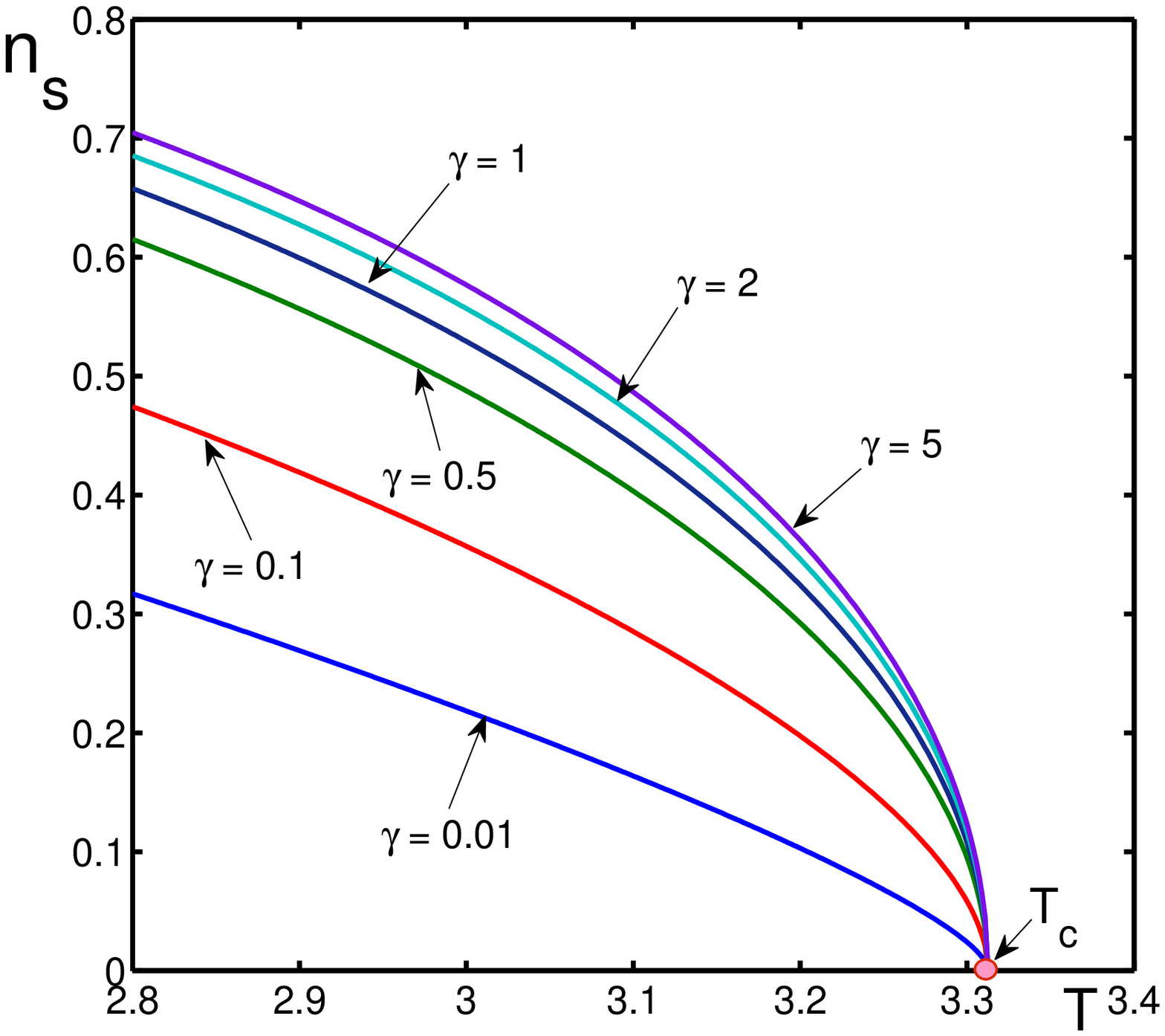} }
\caption{Superfluid fraction $n_s$ as a function of dimensionless temperature 
in the critical region.}
\label{fig:Fig.5}
\end{figure}


\newpage



\begin{thebibliography}{99}

\bibitem{1}
Matsubara T and Matsuda K 1956
{\it Prog. Theor. Phys.} {\bf 16} 569 

\bibitem{2}
Andersen J O 2004
{\it Rev. Mod. Phys.} {\bf 76} 599

\bibitem{3}
Wilson K G and Kogut J 1974
{\it Phys. Rep.} {\bf 12} 75

\bibitem{4}
Ceperley D M 1995
{\it Rev. Mod. Phys.} {\bf 67} 279. 

\bibitem{5}
Boninsegni M, Prokofiev N V and Svistunov B V 2006
{\it Phys. Rev.} E {\bf 74} 036701

\bibitem{6}
Wirth F W and Hallock R B 1987
{\it Phys. Rev.} B {\bf 35} 34 

\bibitem{7}
Pethick C J and Smith H 2008
{\it Bose-Einstein Condensation in Dilute Gases}
(Cambridge: Cambridge University) 

\bibitem{8}
Courteille P W, Bagnato V S and Yukalov V I 2001
{\it Laser Phys.} {\bf 11} 659 

\bibitem{9}
Yukalov V I 2009
{\it Laser Phys.} {\bf 19} 1

\bibitem{10}
Yukalov V I 2011
{\it Phys. Part. Nucl.} {\bf 42} 460 

\bibitem{11}
Luban M 1962
{\it Phys. Rev.} {\bf 128} 965

\bibitem{12}
Luban M and Grobman W D 1966
{\it Phys. Rev. Lett.} {\bf 17} 182

\bibitem{13}
Reatto L and Straley J P 1969
{\it Phys. Rev.} {\bf 183} 321

\bibitem{14}
Kita T 2005
{\it J. Phys. Soc. Jap.} {\bf 74} 1891

\bibitem{15}
Kita T 2006
{\it J. Phys. Soc. Jap.} {\bf 75} 044603  

\bibitem{16}
Baym G and Grinstein G 1977
{\it Phys. Rev.} D {\bf 15} 2897

\bibitem{17}
Olivares-Quiroz L and Romero-Rochin V 2010
{\it J. Phys. B: At. Mol. Opt. Phys.}  {\bf 43} 205302 

\bibitem{18}
Lieb E H, Seiringer R, Solovej J P and Yngvason J 2005
{\it The Mathematics of the Bose Gas and Its Condensation}
(Basel: Birkh\"{a}user)

\bibitem{19}
Yukalov V I 2007
{\it Laser Phys. Lett.} {\bf 4} 632  

\bibitem{20}
Bogolubov N N 1947
{\it J. Phys. (Moscow)} {\bf 11} 23

\bibitem{21}
Bogolubov N N 1947
{\it Moscow Univ. Phys. Bull.} {\bf 7} 43

\bibitem{22}
Shohno N 1964
{\it Prog. Theor. Phys.} {\bf 31} 553 

\bibitem{23}
Chernyak V, Choi S and Mukamel S 2003
{\it Phys. Rev.} A {\bf 67} 053604

\bibitem{24}
Yukalov V I and Yukalova E P 2005
{\it Laser Phys. Lett.} {\bf 2} 506  

\bibitem{25}
Popov V N 1983
{\it Functional Integrals in Quantum Field Theory and Statistical Physics}
(Dordrecht: Reidel)

\bibitem{26}
Popov V N 1987
{\it Functional Integrals and Collective Modes}
(New York: Cambridge University)  

\bibitem{27}
Girardeau M and Arnowitt R 1959
{\it Phys. Rev.} {\bf 113} 755

\bibitem{28}
Girardeau M 1962
{\it J. Math. Phys.} {\bf 3} 131

\bibitem{29}
Morgan S A 2000
{\it J. Phys. B: At. Mol. Opt. Phys.} {\bf 33} 3847
 
\bibitem{30}
Shi H and Griffin A 1998
{\it Phys. Rep.} {\bf 304} 1

\bibitem{31}
Yukalov V I 2008
{\it Ann. Phys.} {\bf 323} 461

\bibitem{32}
Bogolubov N N 1967
{\it Lectures on Quantum Statistics}, Vol. 1 (New York: Gordon and Breach)

\bibitem{33}
Bogolubov N N 1970
{\it Lectures on Quantum Statistics}, Vol. 2 (New York: Gordon and Breach)

\bibitem{34}
Hugenholtz N M and Pines D 1959
{\it Phys. Rev.} {\bf 116} 489 

\bibitem{35}
Ginibre J 1968
{\it Commun. Math. Phys.} {\bf 8} 26

\bibitem{36}
Yukalov V I 2005
{\it Phys. Rev.} E {\bf 72} 066119

\bibitem{37}
Yukalov V I and Yukalova E P 2006
{\it Phys. Rev.} A {\bf 74} 063623

\bibitem{38}
Yukalov V I and Yukalova E P 2007
{\it Phys. Rev.} A {\bf 76} 013602

\bibitem{39}
Hohenberg P C and Martin P C 1965
{\it Ann. Phys.} {\bf 34} 291

\bibitem{40}
Yukalov V I 2006
{\it Laser Phys.} {\bf 16} 511

\bibitem{41}
Yukalov V I 2011
{\it Phys. Lett.} A {\bf 375} 2797

\bibitem{42}
Kalos M H, Levesque D and Verlet L 1974
{\it Phys. Rev.} A {\bf 9} 2178

\bibitem{43}
Giorgini S, Boronat J and Casulleras J 1999
{\it Phys. Rev.} A {\bf 60} 5129 

\bibitem{44}
Rossi M and Salasnich L 2013
{\it Phys. Rev.} A {\bf 88} 053617

\bibitem{45}
Yukalov V I 2013
{\it Laser Phys.} {\bf 23} 062001

\end{thebibliography}
\end{document}